\documentclass[aps,prl,reprint,superscriptaddress,amsfonts,amssymb,amsmath,a4paper,showpacs]{revtex4-1}
\usepackage{bm}
\usepackage[dvipdfm]{graphicx}
\usepackage{dcolumn}

\begin{document}

\title{Critical Bottleneck Size for Jamless Particle Flows in Two Dimensions}

\author{Takumi Masuda}
\affiliation{Department of Aeronautics and Astronautics, Faculty of 
Engineering, University of Tokyo, Hongo, Bunkyo-ku, Tokyo 113-8656, Japan}
\author{Katsuhiro Nishinari}
\affiliation{Department of Aeronautics and Astronautics, Faculty of 
Engineering, University of Tokyo, Hongo, Bunkyo-ku, Tokyo 113-8656, Japan}
\author{Andreas Schadschneider}
\affiliation{Institut f\"{u}r Theoretische Physik, Universit\"{a}t zu 
K\"{o}ln, 50937 K\"{o}ln, Germany}
\date{\today}

\begin{abstract}
  We propose a simple microscopic model for arching phenomena at
    bottlenecks.  The dynamics of particles in front of a bottleneck
  is described by a one-dimensional stochastic cellular automaton on a
  semicircular geometry. The model reproduces oscillation
  phenomena due to formation and collapsing of arches.  It predicts
  the existence of a critical bottleneck size for continuous particle
  flows.  The dependence of the jamming probability on the system size is
  approximated by the Gompertz function. The analytical results
are in good agreement with simulations.
\end{abstract}

\maketitle

Granular materials are many-particle systems that display interesting
and unintuitive physical properties \cite{Mehta,Aronson06}.  One of
their most important types of behavior is formation of arches which leads to a mutual arrest of their constituent particles in front of a bottleneck.
Such situation is usually called a "jam".  Usually it is an
undesirable state since it causes many problems, e.g., in industrial
applications. It often occurs in systems such as traffic \cite{CSS00},
granular flow through a hopper \cite{To01} and escaping stampedes
during evacuations \cite{HelbingFV00}.  Cates et al., in their
comprehensive study \cite{Cates98}, suggest that jammed systems should
be categorized as a new class ''fragile matter", i.e.  materials which
respond to applied stress by reorganizing their internal structures
through force chains.  Liu and Nagel \cite{LiuNagel98} extend the
concept not only to grains, bubbles and droplets but also to glass
transitions.  One focus of recent studies on granular flows has been
on bottleneck flows with external perturbations, e.g., vibrations.
Vibrated granular flows exhibit intermittent behavior, which reflects
phase transitions between a jamming and an unjamming state.

Several experiments have revealed properties of granular flows through
a bottleneck.  The most important one is the existence of a critical
outlet size above which no arches appear \cite{To01}.  However, some
empirical laws do not determine the critical outlet size
\cite{Janda08}.  In addition, the two states of intermittent flows
alternate randomly and lifetime distributions have been investigated.
The avalanche size, defined as the number of grains passing through a
bottleneck during a single unjamming state, follows an exponential
distribution \cite{Janda08,Zuriguel05,HelbingJMJH06,Mankoc09,Janda09}.
On the other hand, the duration of an unjamming state obeys power law
and its expectation value does not converge for low magnitudes of
vibration \cite{Janda09}.
        
In some situations, pedestrian crowds exhibit collective phenomena
similar to those in granular materials, e.g., lane formation as in
oppositely charged colloids \cite{Visser11} and for evacuation flows
at bottlenecks \cite{HelbingFV00,HelbingBJW05}. The latter shows very
similar behavior to a granular flow since also in pedestrian crowds
formation and collapsing of arches has been observed.  Although
granular materials require external perturbations to resume flows,
pedestrian crowds rapidly destroy clogging by self-adjustment.

\begin{figure}[ht]
\includegraphics[width=5cm]{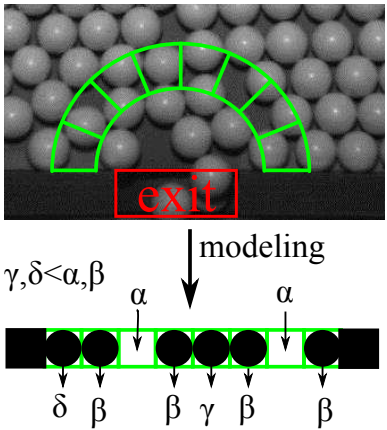}
\caption{Definition of the model.
  Top: A semicircle which is slightly larger than the width of the
  exit is divided into discrete sites which can contain at most one
  particle.  Bottom: Definition of a 1-d stochastic cellular automaton
  characterized by four parameters $\alpha,\beta,\gamma$ and $\delta$.
  An arch corresponds to the configuration where all sites are
  occupied.  The arrow into a site represents a particle "inflow"
  corresponding to particle creation at a rate $\alpha$.  The arrows
  pointing out of sites indicate the "outflow" which is defined by 3-site
  interactions.  In the bulk it occurs with rates $\beta$ or $\gamma$
  and at the boundaries with rate $\beta$ or $\delta$.
\label{discharging}}
\end{figure}

In the following we propose a simple model that captures the essence
of the observed behavior of many-particle systems near a bottleneck,
e.g., oscillation phenomena.  Although particle flows usually are
three-dimensional, we focus here on two-dimensional realizations which are
relevant for pedestrian dynamics, but have also been studied for
granular materials.  
For simplicity, we ignore fluctuations that occur in
the bulk of granular assemblies \cite{HelbingJMJH06}.
Instead, we focus on properties of intermittent behavior which stem 
from arching phenomena. The precise structure of the arches is
not relevant for the properties of the flow.
This assumption allows us to formulate the dynamics of the particles 
by a one-dimensional stochastic cellular automaton.
Its sites are arranged in a semicircular shape which reflects the typical 
form of arches (Fig.~\ref{discharging}). 
Here we have assumed that no
arches appear in the area nearer to the bottleneck than the semicircle, which implies that its size is of the order of the bottleneck width.
If the site size is chosen as the typical size of the particles
(grains), each site can be occupied by at most one particle. Hence,
each site $j$ can be in two different states, empty ($s_j=0$) or
occupied ($s_j=1$).  The configuration $(1,\ldots,1)$ where all sites
are occupied represents arch formation.  If $P(C)$ denotes the
probability of finding a configuration $C=(s_1,\ldots,s_L)$ in the
steady state, the arching probability is given by
$P_\mathrm{arch}=P(1,\ldots,1)$.

In order to define the dynamics of the model we assume that the bulk
of the granular assembly acts as a particle bath which supplies
particles to the system at a constant rate $\alpha$.
Then empty sites
become occupied with the probability of $\alpha$ which can be
interpreted as the probability that a particle finds an available gap.
It is called ''inflow" in the following.  The ''outflow" is
represented by the annihilation of a particle.  The probability of
this process depends on the occupancy of the two neighboring sites.
If both are occupied, then the particle is annihilated with
probability $\gamma$.  For the other cases the outflow probability is
$\beta$.  At the boundary sites, the "outflow" depends only on a
single neighboring site.  It occurs with a probability of $\delta$
when the site is occupied and with $\beta$ for an empty neighbor site.
In the physical regime, $\gamma$ and $\delta$ are smaller than
$\beta$, since these parameters capture the effects of friction among
grains and walls.  Hence $\gamma$ and $\delta$ decrease as friction
becomes stronger.  In each step, these update rules are applied to a
randomly chosen site (random-sequential update), which is an
approximate realization of a stochastic process in continuous time.

A flow rate $Q(C)$ for a configuration $C$ can be defined as the
probability that an outflow event occurs.  In particular, the flow
rate for the arching configuration
$Q(1,\ldots,1)=(2\delta+(L-2)\gamma)/L=:Q_\mathrm{arch}$
indicates the probability that an arch breaks.  Hence the lifetime
distribution of arches is given by
$Q_\mathrm{arch}(1-Q_\mathrm{arch})^{t-1}$ which has the expectation
value $1/Q_\mathrm{arch}$.  In our model, arches are not stable in the
sense that they have an infinite lifetime. We therefore introduce a
stability threshold $N$ and consider all arches with lifetimes larger
than $N$ as "stable".  Then an arch is stable with probability
$S:=1-Q_\mathrm{arch}\sum^N_{t=1}(1-Q_\mathrm{arch})^{t-1}
=(1-Q_\mathrm{arch})^N$.  The lifetime distribution of stable arches
($t>N)$ is given by $Q_\mathrm{arch}(1-Q_\mathrm{arch})^{t-(N+1)}$
which has the expectation value $1/Q_\mathrm{arch} + N$.

For simplicity, we restrict our attention to the cases where $\alpha =
\beta \neq 0$ and $\gamma=\delta$. The first condition implies that
inflow and outflow rates are identical when no friction acts.  The
second identity implies that the friction between particles and
between particles and walls are identical. In this situation,
$Q_\mathrm{arch}$ is independent of the system size $L$. We introduce
a new parameter $\varepsilon=\gamma/\alpha$ so that $0\leq
  \varepsilon \leq 1$ in the physical regime. We first consider two
limiting cases.  In the case $\varepsilon=0$, flow cannot
resume once an arch has formed.  This situation corresponds to an
absorbing state where the system attains a trivial stationary
  state without dynamics.  Similar behavior is observed when granular
materials flow through a narrow hopper without vibration.  When
$\varepsilon =1$, all configurations appear uniformly in the steady
state since inflow and outflow occur at the same rate.
Therefore the probability for each configuration is $1/2^L$.
We can interpret the parameter $\varepsilon$ as an indicator 
for the magnitude of destabilization of arches since the 
conditions $\varepsilon=0$ and $\varepsilon=1$ correspond to 
jamming and continuous flow, respectively. Additionally,
$\varepsilon$ accounts for arch destabilization by pedestrians. 
Consider a situation where arches are formed during
a rush through a bottleneck.  
Because of the high velocity of the pedestrians and the large
friction between them this situation is described by large
values of $\alpha$ and $\beta$ and small values of $\gamma$ 
and $\delta$.  As a consequence, $\varepsilon$ is small and
can be viewed as an indicator for the pedestrian's discipline 
near the exit.

\begin{figure}[ht]
\includegraphics[width=5.5cm]{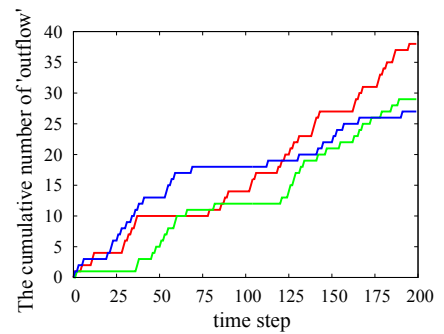}
\caption{
  The dynamical behavior of the model for
  $\alpha=\beta=0.9,\gamma=\delta=0.1,L=3$ and different realizations
  of the stochastic dynamics.  The vertical axis indicates the
  cumulative number of outflowing (annihilated) particles.
\label{simulation}}
\end{figure}

The collective behavior observed in simulations is in good qualitative
agreement with experiments on granular materials.  The dynamical
behavior of the model indicates the presence of two states: jamming
and continuous flow.  Jamming is represented in the graph
(Fig.~\ref{simulation}) by horizontal regions, where due to the
existence of an arch no particles are annihilated.  The other parts
show nonvanishing particle flows.  A similar intermittent behavior
with random alternation between two such states can be observed in
granular flows and escaping stampedes \cite{HelbingBJW05,Janda09}.

Let us now focus on the avalanche size $m$.  In our model, the avalanche
size is defined as the number of outflowing particles between two
successive "stable" arches.  Presuming that avalanche sizes are
distributed exponentially as observed in experiments, we are
interested in their expectation value alone.  It is obtained by
dividing the number of outflowing particles per unit time by the
number of avalanches.  The latter is identical to the number of 
stable arches since they occur alternately.  Therefore, it is given by
$SP_\mathrm{arch}Q_\mathrm{arch}$ where
$P_\mathrm{arch}Q_\mathrm{arch}=P_\mathrm{arch}/(1/Q_\mathrm{arch})$
is the number of arches per unit time.  In addition, the number of
outflowing particles per unit time is obtained as weighted average
of flow rates for configurations on their distribution.  Introducing
$2^L$-dimensional vectors $|P\rangle$ and $\langle Q|$ such that
\begin{align}
|P\rangle =& \sum_{(s_1,\ldots,s_L)} 
P(s_1,\ldots,s_L)|s_1,\ldots,s_L\rangle, \\
\langle Q| =& \sum_{(s_1,\ldots,s_L)} 
Q(s_1,\ldots,s_L)\langle s_1,\ldots,s_L|
\end{align}
where
\begin{align}
|s_1,\cdots,s_L\rangle = |s_1\rangle \otimes\cdots\otimes|s_L\rangle, \;
|0\rangle = \begin{pmatrix} 1 \\ 0 \end{pmatrix}, \;
|1\rangle = \begin{pmatrix} 0 \\ 1 \end{pmatrix}, \notag
\end{align}
we can write the weighted average as $\langle Q|P\rangle$ .
The summation $\sum\nolimits_{(s_1,\ldots,s_L)}$ is over all configurations.
Thus the expectation value of avalanche sizes $m$ is represented as
\begin{align}
m =& \frac{1}{S R(\varepsilon,L)} &
\mathrm{where\ \ } R(\varepsilon,L) =& \frac{Q_\mathrm{arch}
P_\mathrm{arch}}{\langle Q|P\rangle}.
\label{ExpectationAvalancheSize}
\end{align}
The form of (\ref{ExpectationAvalancheSize}) implies that the
  variables $(\gamma,\varepsilon,L,N)$ of $m$ are separated so that
  $R(\varepsilon,L)$ depends only on physical properties of the system
  and $S$ contains parameters $(\gamma,N)$ which do not have a simple
  interpretation in real systems. Since $(\gamma,N)$ depend on the
length of the time step they have to be determined empirically
for each experiment.

In the following, we consider the distribution of configurations in
the steady state $|P\rangle$ to represent (\ref{ExpectationAvalancheSize})
in an explicit form.  Its time evolution is given by the master
equation.  Using the quantum formalism (see e.g.,
\cite{Schuetz,SCN10}), it can be cast in the form of a Schr\"{o}dinger
equation with some ''Hamiltonian" $H$ defined by the transition rates.
In the stationary state it takes the form
\begin{align}
H|P\rangle = 0\,.
\label{master_equation}
\end{align}
The Hamiltonian is readily constructed from the update rule of the model.  
Because of the 3-site interaction the Hamiltonian of our model is more
complicated than e.g., the asymmetric exclusion process.

We readily deduce $\det H=0$ since the master equation implies that
$H$ has an eigenvalue 0.  From the general relation
$H(\mathrm{adj}H)|v\rangle =(\det H)|v\rangle=0$, where $|v\rangle$ is
an arbitrary vector, it follows that the formal solution of
(\ref{master_equation}) is $(\mathrm{adj}H)|v\rangle$.  We choose
$|v\rangle$ as the vector 
$|V\rangle=\sum_{(s_1,\ldots,s_L)} |s_1,\ldots,s_L\rangle$.
We will show elsewhere that the choice of $|v\rangle$ does not depend on 
the form of $H$.
$|P\rangle$ is given by
\begin{align}
|P\rangle = \frac{(\mathrm{adj\,}H)|V\rangle}{\langle
V|(\mathrm{adj\,}H)|V\rangle}.
\end{align}
The denominator of $|P\rangle$ is the normalization constant for the
conservation of probabilities. 
After a cumbersome calculation, we obtain a simpler form of 
$\langle w|P\rangle$ where $\langle w|$ is an arbitrary vector:
\begin{align}
\langle w|P\rangle 
&= \frac{\mathrm{det}[\; H+|V\rangle\langle w|\;]}{\mathrm{det}[\; 
H+|V\rangle\langle V| \;]}.
\label{ProbFormula}
\end{align}
By using (\ref{ProbFormula}),
$R(\varepsilon,L)$ is given by
\begin{align}
R(\varepsilon,L)
&= \frac{\mathrm{det}[\; H+|V\rangle\langle 1,\cdots,1|\;]}{\mathrm{det}
[\; H+|V\rangle\langle Q|/Q_\mathrm{arch} \;]}.
\label{RFormula}
\end{align}
We emphasize that the result (\ref{ProbFormula}) 
is exact and holds for any stochastic cellular automaton model 
with finite number of sites.

\begin{figure}[ht]
\includegraphics[width=5.5cm]{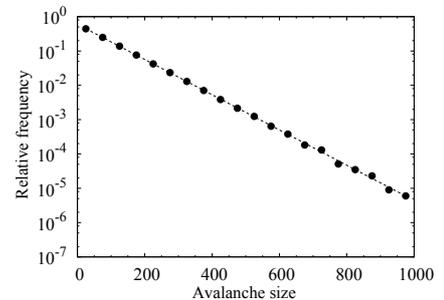}
\caption{
  Histogram of avalanche sizes.  The dotted line is calculated with
  (\ref{ExpectationAvalancheSize}) under the presumption that the
  distribution is exponential.  Dots are simulation results
  for $\alpha=\beta=0.7$, $\gamma=\delta=0.3$, $L=4$, and $N=10$.
\label{avalanche}}
\end{figure}

As shown in Fig.~\ref{avalanche}, the simulation results agree well with
the presumption that avalanche sizes in our model are distributed
exponentially.  The exponential distribution of avalanche sizes has
also been observed in experiments and other simulations of granular
flow \cite{Janda09,Zuriguel05,Janda08}.

Let us now consider the jamming probability $J$.  It is interpreted in our
model as the probability that an avalanche size is less than a
threshold $M$.  Hence, it is obtained by integrating the avalanche
size distribution from 0 to $M$:
\begin{align}
J =& 1-\exp(-M/m) = 1-\exp[-SMR(\varepsilon,L)].
\end{align}
\begin{figure}[ht]
\includegraphics[width=5.5cm]{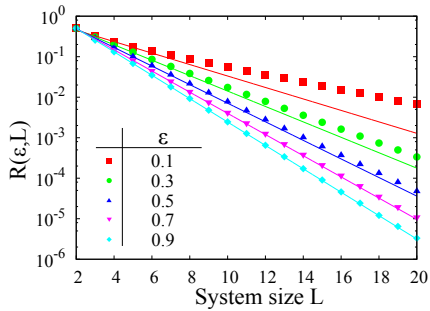}
\caption{
  Dependence of $R(\varepsilon,L)$ on $L$.  The dots correspond to
  simulation results for different values of $\varepsilon$.  The lines
  are fixed by the two points $R(\varepsilon,3)$ and
  $R(\varepsilon,4)$ for corresponding $\varepsilon$.  It is found
  that for $\varepsilon\geq 0.5$. $R(\varepsilon,L)$ can be
  approximated by an exponential function.  }
\label{RatioGraphColor}
\end{figure}

Although the dependence of $R(\varepsilon,L)$ on $\varepsilon$ has a
rational form as implied from (\ref{RFormula}), the dependence on $L$
is nontrivial.  This fact motivates us to approximate
$R(\varepsilon,L)$ by an analytical function.
Figure.~\ref{RatioGraphColor} shows that $R(\varepsilon,L)$ is
represented by an exponential function
$A(\varepsilon)\exp[-B(\varepsilon)L]$ for $\varepsilon\geq 0.5$.  In
fact, this assumption can be justified for the case $\varepsilon=1$.
Identifying $A(\varepsilon)$ and $B(\varepsilon)$ with
$R(\varepsilon,3)$ and $R(\varepsilon,4)$, we can write the jamming
probability with the Gompertz function as
\begin{align}
J(\varepsilon,L) =& 1-\exp[-A(\varepsilon)SM\exp[-B(\varepsilon)L]\;], 
\label{JammingEquation} \\
A(\varepsilon) =& R(\varepsilon,3)^4R(\varepsilon,4)^{-3}, \\
B(\varepsilon) =& \log R(\varepsilon,3)-\log R(\varepsilon,4).
\end{align}
$R(\varepsilon,3)$ and $R(\varepsilon,4)$ are calculated from
(\ref{RFormula}) as
\begin{align}
R(\varepsilon,3) =& \frac{\varepsilon+23}{4(7\varepsilon+17)}, \\
R(\varepsilon,4) =& \frac{11\varepsilon^2+78\varepsilon+103}{2
(24\varepsilon^3+181\varepsilon^2+366\varepsilon+197)}.
\end{align}

\begin{figure}[ht]
\includegraphics[width=5.5cm]{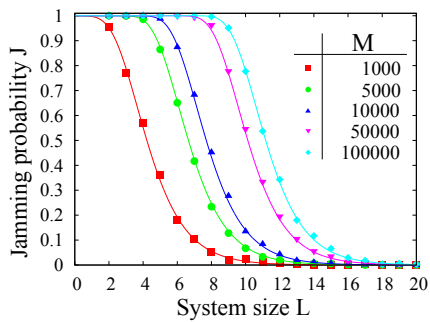}
\caption{Jamming probabilities as functions of system size.
  The plots are simulation results and the lines are defined by
  (\ref{JammingEquation}).  The jamming probabilities gradually
  decrease with increasing system size.  They practically become zero
  already for relatively small system size.  The system parameters are
  $\alpha=\beta=0.45$, $\gamma=\delta=0.4$ and $N=10$.}
\label{JammingGraphColor}
\end{figure}

The simulation results shown in Fig.~\ref{JammingGraphColor} agree well
with our previous assumptions that the avalanche size distribution and
$R(\varepsilon,L)$ are exponential.

The jamming probability $J$ converges to 1 for any system size in the
limit $M\to\infty$ in principle, as deduced from
(\ref{JammingEquation}).  However, at a finite $M$ the jamming
probability becomes 0 at a finite system size $L$ in practice.
In experiments, this fact corresponds to the existence of a critical
outlet size above which no arches appear \cite{To01,Janda08}.

A typical value of $\varepsilon$ may be estimated from experimental
results.  In \cite{Mankoc09}, Mankoc et al.\ introduced the bivariate
model characterized by $p$ and $q$, which indicate the probability
that a particle passes through the outlet without forming an arch and
the probability that a particle is delivered from an arch
respectively.  The parameters have been experimentally estimated as
$p=0.981,q=0.836$ for an outlet of $3.02$ grain diameters width.
Although their experiments are in three- dimensions, we assume that the
results are appropriate for our model.  From the
definition, $q$ can be interpreted in our model as $S=1-q$.
Comparing the
expectation values of avalanche sizes deduced by both models, we
obtain that $R(\varepsilon,L) = (1-p)/(p+q-pq)$.  Additionally, we use
$L\simeq 6.9$ which is reported from experiments in \cite{Garcimartin10} as the number
of particles involved in an arch for the outlet of $3.03$ grains
diameter width.  Then we obtain $\varepsilon\simeq 0.92$.
We interpret the dynamical behavior of particles in front of a
bottleneck as the cellular automaton model with 3-site interactions
arranged in a semicircular shape.  From the simulations and the
analytical results we can conclude that the model reproduces the
  generic behavior which characterizes bottleneck flows in
many-particle systems.  The resulting dynamics exhibits two clear
regions: jamming and continuous flow.  The avalanche size distribution
is exponential and the jamming probability is well approximated by the
Gompertz function.
The expectation value of avalanche sizes and the coefficients of the 
Gompertz function can be determined analytically.
The model reveals the existence of a critical outlet size above which
no arches appear in practice.  The parameter $\varepsilon$, which
characterizes the physical properties of the model, can be estimated by
methods which have been used in previous studies.

The model can be extended to be more compatible with actual particle
flows.  Although we focus on two-dimensional flows for simplicity, the
model can be extended to three-dimensional flows in a straightforward way.
Moreover, we have formulated the model assuming that an arch appears
only in a single semicircular layer.  Again the model can be made more
realistic by considering multiple layers to take into account the effects
of the upstream and allow for variations in arch size.


\end{document}